\documentclass[5p,twocolumn]{elsarticle}
\usepackage{graphicx,epstopdf,titlesec,siunitx,amsmath,fixltx2e,subcaption,textgreek,amsmath,lineno}
\journal{Icarus}
\biboptions{authoryear}

\begin{document}

\begin{frontmatter}

\title{Cloud structure and composition of Jupiter's troposphere from 5-\textmu m Cassini VIMS spectroscopy}

\author[oxford]{R.S. Giles\corref{cor1}}
\ead{giles@atm.ox.ac.uk}
\author[oxford]{L.N. Fletcher}
\author[oxford]{P.G.J. Irwin}

\address[oxford]{Atmospheric, Oceanic \& Planetary Physics, Department of Physics, University of Oxford, Clarendon Laboratory, Parks Road, Oxford, OX1 3PU, UK}

\cortext[cor1]{Corresponding author}

\begin{abstract}

Jupiter's tropospheric composition and cloud structure are studied using Cassini VIMS 4.5-\SI{5.1}{\micro\meter} thermal emission spectra from the 2000-2001 flyby. We make use of both nadir and limb darkening observations on the planet's nightside, and compare these with dayside observations. Although there is significant spatial variability in the 5-\textmu m brightness temperatures, the shape of the spectra remain very similar across the planet, suggesting the presence of a spectrally-flat, spatially inhomogeneous cloud deck. We find that a simple cloud model consisting of a single, compact cloud is able to reproduce both nightside and dayside spectra, subject to the following constraints: (i) the cloud base is located at pressures of~\SI{1.2}{\bar} or lower; (ii) the cloud particles are highly scattering; (iii) the cloud is sufficiently spectrally flat. Using this cloud model, we search for global variability in the cloud opacity and the phosphine deep volume mixing ratio. We find that the vast majority of the 5-\textmu m inhomogeneity can be accounted for by variations in the thickness of the cloud decks, with huge differences between the cloudy zones and the relatively cloud-free belts. The relatively low spectral resolution of VIMS limits reliable retrievals of gaseous species, but some evidence is found for an enhancement in the abundance of phosphine at high latitudes.

\end{abstract}

\begin{keyword}
Jupiter  \sep Atmospheres, composition \sep Atmospheres, structure
\end{keyword}

\end{frontmatter}


\section{Introduction}

The 5-\textmu m atmospheric window is a unique region of Jupiter's spectrum, where a dearth of opacity from the planet's principal infrared absorbers gives us access to parts of the atmosphere that are otherwise hidden from view. The sensitivity of the 4.5-\SI{5.2}{\micro\meter} spectrum peaks in the 4-\SI{8}{\bar} region of Jupiter's troposphere, beneath the planet's topmost cloud decks (Figure~\ref{fig:weighting}). The observed brightness temperatures are therefore highly dependent on the properties of these clouds: the observed radiance varies significantly from the warm, cloud-free belts to the cooler, cloudier zones, a phenomenon first described by~\citet{westphal69}. This sensitivity makes the 5-\textmu m spectral region extremely useful in analysing both the 
composition and cloud structure of Jupiter's middle troposphere.

\begin{figure}[h]
\centering
\begin{subfigure}{0.42\textwidth}
\centering
\includegraphics[width=1.0\textwidth]{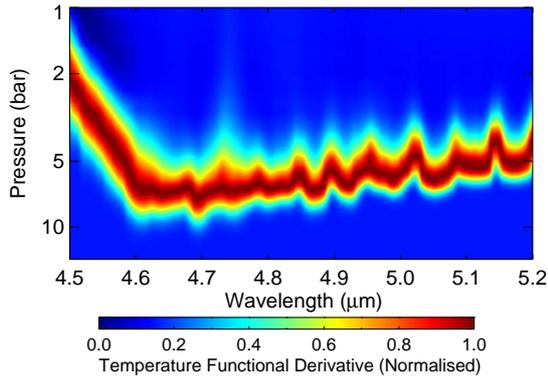}
\caption{The temperature functional derivative for a cloud-free atmosphere, normalised at each wavelength.}
\label{fig:weightinga}
\end{subfigure}
\begin{subfigure}{0.42\textwidth}
\centering
\includegraphics[width=1.0\textwidth]{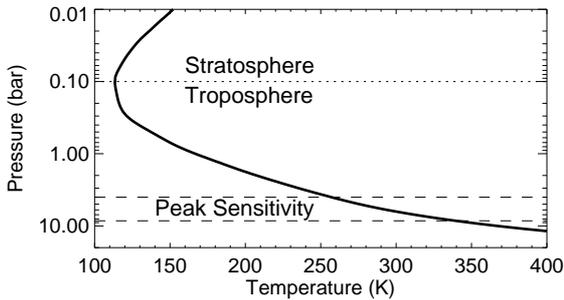}
\caption{Jupiter's temperature-pressure profile.}
\label{fig:weighting}
\end{subfigure}
\caption{The sensitivity of the 5-\textmu m window as a function of wavelength. The peak sensitivity is in the 4-\SI{8}{\bar} region of the troposphere, beneath the planet's topmost cloud decks.}
\label{fig:weighting}
\end{figure}

The  5-\textmu m data used in this study were obtained using the Visual and Infrared Mapping Spectrometer (VIMS,~\citealp{brown04}) on the Cassini spacecraft. In late 2000 and early 2001, Cassini performed a fly-by of Jupiter, reaching a closest approach of 9.7 million km (137 Jovian radii) on December 30 2000. During this period, VIMS made spatially-resolved spectroscopic measurements covering the 0.35-\SI{5.1}{\micro\meter} wavelength range. \citet{sromovsky10b} used this dataset to study features in reflected sunlight at \SI{3}{\micro\meter}, but the  5-\textmu m segment of this dataset has not yet been analysed. In this paper, we use 5-\textmu m VIMS spectra with a range of phase and emission angles to investigate Jupiter's tropospheric composition and cloud structure; we will draw conclusions about the cloud locations and scattering parameters, and we will assess the degree of constraint offered by Cassini/VIMS on gaseous variability. This builds on previous analyses using  5-\textmu m data from Galileo NIMS and Voyager IRIS. The VIMS data gives us full global coverage of Jupiter, allowing us to compare and contrast the 5-\textmu m spectra from the belts and the zones. Additionally, VIMS made observations on Jupiter's nightside and dayside within the space of a few weeks, giving us the opportunity to study the effect of reflected sunlight. 

A typical theoretical cloud condensation model from~\citet{atreya97} produces four cloud levels: NH\textsubscript{3}-ice, NH\textsubscript{4}SH, H\textsubscript{2}O-ice and aqueous-ammonia solution. If a solar chemical composition is assumed, the bases of these cloud levels are located at \SI{0.7}{\bar}, \SI{2.2}{\bar}, \SI{5.4}{\bar} and \SI{5.7}{\bar} respectively. Although models such as this provide valuable insights into the approximate condensation levels for each species, we do not necessarily expect to see the clouds form at these precise locations. The models make assumptions about both the bulk composition of the atmosphere and the production of photochemical products. They also do not account for the complex dynamics of the troposphere, neglecting the vertical mixing which brings up material from below and rains down material from above. All of these factors will act to significantly alter the vertical cloud structure. Observations are therefore fundamental to understanding Jupiter's tropospheric cloud decks. In the following paragraphs, we will briefly summarise the conclusions that have been drawn from previous observational studies.

There are many studies supporting the existence of the uppermost NH\textsubscript{3}-ice cloud deck. 5- and 45-\textmu m observations from Voyager IRIS were analysed by~\citet{gierasch86}, who concluded that there was a highly variable cloud deck at $\sim$\SI{700}{\milli\bar}, near the expected ammonia condensation level. Imaging data from the Galileo spacecraft showed evidence for thick, variable clouds in the 750$\pm$\SI{200}{\milli\bar} region~\citep{banfield98}, again consistent with NH\textsubscript{3}. Further support came in the form of spectroscopic identification of ammonia-ice features, using the 1-\SI{3}{\micro\meter} region of the Galileo NIMS spectra~\citep{baines02} and \SI{9}{\micro\meter} data from Cassini CIRS~\citep{wong04}. ~\citet{baines02} found that their ammonia-ice spectroscopic signatures were present in less than 1\% of Jupiter's clouds, but it has been suggested that this may be due to a hydrocarbon haze coating the ammonia--ice particles and masking the absorption features~\citep{kalogerakis08}.

There are additional studies which support the existence of both an upper NH\textsubscript{3} cloud and a lower cloud formed of NH\textsubscript{4}SH. ~\citet{matcheva05} found clouds in the 900-\SI{1100}{\milli\bar} region from Cassini CIRS observations at  \SI{7.18}{\micro\meter}, which they concluded were probably composed of an upper layer of NH\textsubscript{3} and a lower layer of NH\textsubscript{4}SH. \citet{sromovsky10b} analysed 3-\textmu m data from Cassini VIMS, and found that a \SI{500}{\milli\bar} cloud composed of both NH\textsubscript{3} and NH\textsubscript{4}SH provided the best fit; they suggest that rapid upwelling could lead to the presence of NH\textsubscript{4}SH particles well above the expected condensation point. Further evidence for a NH\textsubscript{4}SH cloud comes from analyses of the 5-\textmu m Galileo NIMS data. \citet{irwin01} performed retrievals on the NIMS data and determined that the cloud that provides the majority of the \SI{5}{\micro\meter} opacity is located at around \SI{1.5}{\bar}. This was supported by~\citet{irwin02}, who performed a principal component analysis on the same data and once again found that the dominant opacity variations occur in the 1-\SI{2}{\bar} region. Based on the location, both studies concluded that the dominant cloud was likely to be composed of ammonium hydrosulfide. 

Because of its predicted location deep in the lower troposphere, the water cloud has been difficult to observe. \citet{banfield98} used Galileo SSI observations to identify a deep cloud (located at a pressure greater than \SI{4}{\bar}) in one region close to the Great Red Spot, which they concluded was likely to be composed of water. Similarly, ~\citet{simon-miller00} found evidence for the cloud in very small regions of the planet by identifying a water ice feature near \SI{44}{\micro\meter} in Voyager IRIS spectra. This limited detection may be due to the fact that the water cloud is only visible in small regions where the rest of the atmosphere is particularly cloud-free.

In addition to studying the cloud structure, the 5-\textmu m region allows us to retrieve abundances for tropospheric species and determine if there is any global variability. Disequlibrium species, such as PH\textsubscript{3}, are expected to have higher mixing ratios in regions of upwelling, as the gas is brought up from deeper pressures where it is more abundant. \citet{irwin04} and~\citet{fletcher09} find that Cassini CIRS data shows a PH\textsubscript{3} enhancement at the equator compared to the belts on either side, supporting this view. Similarly, zones are expected to have a higher water abundance than belts, as moist air is brought up from below; ~\citet{carlson92}, using Voyager IRIS data, found that low relative humidities were correlated with bright  5-\textmu m regions, so the bright belts were found to be depleted in water.

In this paper, we investigate the constraints placed on Jupiter's tropospheric clouds by  5-\textmu m VIMS data, and search for any spatial variability in Jupiter's tropospheric gaseous species. We find that the VIMS observations can be modelled using a single, highly reflecting cloud layer located at a pressure of \SI{1.2}{\bar} or lower, and that this cloud is responsible for the vast majority of the global spectral variability. The relatively low spectral resolution of VIMS limits the gaseous retrievals, but we find some evidence for an enhancement in the abundance phosphine at high latitudes.

\section{Observations}
\label{sec:observations}

\subsection{VIMS data and calibration}

Between October 2000 and March 2001, tens of thousands of images of the Jovian system were taken by the Cassini spacecraft. These included  5\textmu m measurements taken by the Visual and Infrared Mapping Spectrometer (VIMS,~\citealp{brown04}). VIMS is made up of two imaging spectrometers, one covering the visible spectral range (VIMS-V) and the other covering the infrared (VIMS-IR). The present study uses data from the 4.5-\SI{5.1}{\micro\meter} region from VIMS-IR.

VIMS-IR covers the wavelength range 0.85-\SI{5.1}{\micro\metre} with a spectral sampling of \SI{16.6}{\nano\metre}, giving 256 wavelength channels. The spectral resolution varies slightly across the wavelength range, and has an average value of \SI{18.7}{\nano\meter} in the 5-\textmu m window (K. Baines, personal communication). The instantaneous field of view of VIMS-IR is 0.25$\times$\SI{0.50}{\milli\radian}. By using a two-axis scan mechanism, the instantaneous field of view is stepped across the scene to scan a full 64$\times$64 pixel image. The total field of view is 32$\times$\SI{32}{\milli\radian}, which is several times larger than the size of Jupiter's disk at the distance of closest approach (Jupiter has a diameter of \SI{15}{\milli\radian} at distance of 9.7 million km).

The VIMS cubes were processed using ISIS3 (Integrated Software for Imagers and Spectrometers), a software package provided by the USGS~\citep{anderson04}. The VIMS Science Team calibration~\citep{mccord04} is included in this software and geometries are assigned to the cubes using NASA/JPL SPICE kernels, so both radiometric and geometric calibration are achieved. The final result is a data cube of dimensions 64$\times$64$\times$256 (one spectral and two spatial dimensions) and a backplane cube of dimensions 64$\times$64$\times$13, including information such as the latitude, longitude, phase angle and emission angle of each spatial point.

\subsection{Data selection}
\label{sec:data_selection}

VIMS observations were made both on the dayside of the planet as the spacecraft approached in late 2000 and on the planet's nightside as the spacecraft moved away from Jupiter in early 2001. Nightside observations simply show the thermal emission from the planet, whereas the dayside observations also include a component from reflected sunlight which further complicates the analysis; Sections~\ref{sec:basic},~\ref{sec:cloud_constraints} and~\ref{sec:variations} focus on the nightside data, and Section~\ref{sec:dayside} compares these results to the dayside observations.

For the nightside, a set of 30 data cubes from a 7 hour period on 11 January 2001 were chosen, when Cassini was a distance of 15 million km from Jupiter. The use of multiple cubes vastly increases the number of available data points and allows us to cover more longitudes, while the fairly limited time period means that atmospheric conditions are unlikely to have changed. The date was chosen so as to avoid the day-night boundary (to obtain purely nightside emission), while maximising the spatial resolution. For the subsequent dayside analysis, a set of data cubes from a 10 hour period on 17 December 2000 were used. These observations were made at a similar distance from the planet as the 11 January ones (16 million km), giving a comparable spatial resolution of $\sim$\SI{8000}{\kilo\meter} ($\sim$ half the size of the GRS).

\begin{figure}
\centering
\begin{subfigure}{0.5\textwidth}
\centering
\includegraphics[width=0.95\textwidth]{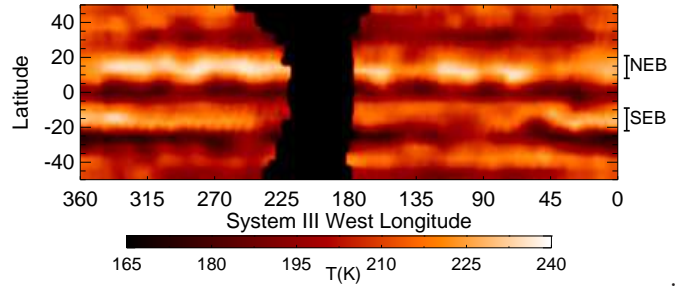}. 
\caption{Nightside data from 11 January 2001.}
\label{fig:vimsmap_night}
\end{subfigure}
\begin{subfigure}{0.5\textwidth}
\centering
\includegraphics[width=0.95\textwidth]{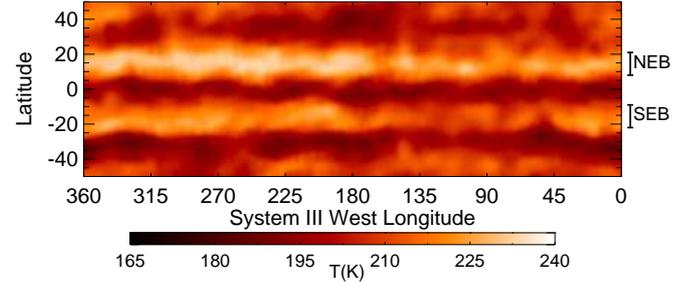}
\caption{Dayside data from 17 December 2000.}
\end{subfigure}
\caption{Brightness temperature maps from \SI{5.0}{\micro\meter} VIMS data on two dates, giving both nightside and dayside observations of Jupiter. Smoothing has been applied to these maps. Black regions correspond to segments where no data was obtained. The GRS is located at $\sim50^{\circ}$W and did not move significantly between the two datasets.}
\label{fig:vimsmap}
\end{figure}

Cylindrical projections of both the nightside and the dayside datasets are shown in Figure~\ref{fig:vimsmap}. In both panels, the difference between the warm belts and the cooler zones can be seen. In the nightside observations, there is a $\sim$\SI{70}{\kelvin} brightness temperature difference between the warmest and the coolest regions of the planet. For the dayside observations, this contrast is reduced to $\sim$\SI{50}{\kelvin}; the cloudy zones reflect more light than the relatively cloud-free belts, meaning that the presence of sunlight has a greater impact on the cool zones than on the warm belts. Zonal mean radiances were then extracted from both the nightside and the dayside datasets. The spectra were divided into forty-one latitudinal bins covering the range $-50^{\circ}$ to $+50^{\circ}$ at $2.5^{\circ}$ intervals, each with a $5^{\circ}$ width. For each latitudinal bin, all spectra with an emission angle within $10^{\circ}$ of the minimum value were averaged to produce one final nadir spectrum for each latitude segment.

Following ~\citet{fletcher11}, the final errors assigned to these averaged VIMS spectra were 12\% of the mean radiance in the 4.5-\SI{5.2}{\micro\meter} range. This is a conservative value, including both quadrature-estimated errors due to pre-flight calibration and forward-model uncertainties on spectral line data. The constant error margin throughout the window prevents retrievals being weighted towards low-radiance regions. The size of the VIMS errors is further discussed in Section~\ref{sec:variations}.

\begin{figure}
\centering
\begin{subfigure}{0.5\textwidth}
\centering
\includegraphics[width=0.95\textwidth]{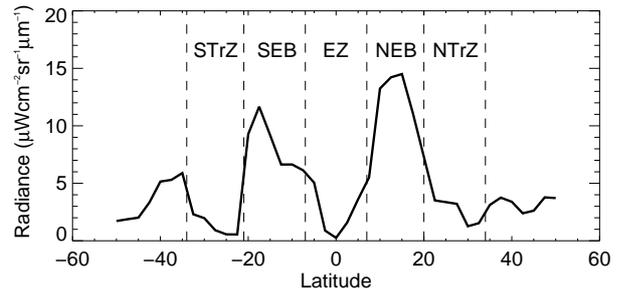}. 
\caption{\SI{5.0}{\micro\meter} radiance as a function of latitude.}
\end{subfigure}
\begin{subfigure}{0.5\textwidth}
\centering
\includegraphics[width=0.95\textwidth]{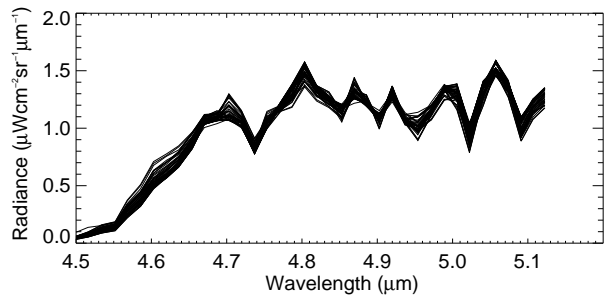}
\caption{The normalised spectrum for each latitudinal bin.}
\end{subfigure}
\caption{VIMS nightside spectra, divided into $5^{\circ}$ wide latitudinal bins and zonally averaged. The upper panel shows how the absolute radiance at \SI{5.0}{\micro\meter} varies with latitude. The lower panel shows the full spectrum for each latitudinal bin, normalised to allow a comparison of their shapes.}
\label{fig:spectra}
\end{figure}

The results of this process for the nightside data can be seen in Figure~\ref{fig:spectra}. Figure~\ref{fig:spectra}b shows the forty-one spectra from the different latitudes, all normalised to $1.0$ to allow a comparison of spectral shapes, while Figure~\ref{fig:spectra}a shows the absolute radiances at \SI{5.0}{\micro\meter} as a function of latitude. While Figure~\ref{fig:spectra}a shows that there is a factor of 50 difference between the hottest and coldest spectra, Figure~\ref{fig:spectra}b shows that the shape of the spectra remains almost identical throughout. This cannot be explained by variations in the chemical composition or temperature profile of the atmosphere. The only parameter capable of reproducing the huge variability in the radiance while retaining the same spectral shape is the cloud optical thickness, as we show quantitatively in the following sections.

\section{Spectral modelling}

\subsection{Radiative transfer model and retrieval algorithm}
\label{sec:radiativetransfer}

The Jupiter VIMS spectra were analysed using the NEMESIS software developed by~\citet{irwin08}. This software has previously been used to analyse the 5-\textmu m VIMS spectra of Saturn~\citep{fletcher11} and an earlier version was used to analyse Jupiter NIMS data~\citep{irwin98,irwin01,nixon01}. NEMESIS is made up of a radiative transfer code that computes the emergent radiation for a given atmospheric profile and an optimal estimation retrieval algorithm which iteratively determines the best-fit atmospheric parameters for an observed spectrum.

The radiative transfer code solves the equation of radiative transfer to calculate the top-of-atmosphere radiation as a function of wavelength. The input information required includes a priori estimates of the atmospheric composition and the temperature profile, as well as absorption line information for each molecule. Rather than computing each individual absorption line, NEMESIS uses the correlated-k approximation~\citep{lacis91} to reduce the computational time. This approximation allows us to reshuffle the absorption coefficients within a small spectral interval into ascending order, producing a smoothly varying function that requires fewer steps in numerical integration~\citep{irwin08}. NEMESIS has the capability to model a multiple-scattering atmosphere. The full Mie-scattering phase function is approximated using a combined Henyey-Greenstein function~\citep{henyey41} and the calculations are performed using a matrix-operator approach~\citep{plass73}. 

The NEMESIS retrieval algorithm determines the atmospheric parameters from an observed input spectrum through an iterative process. Starting with an initial a priori atmospheric profile, synthetic ``forward-models'' are generated using the radiative transfer code. These synthetic spectra are compared to the observed spectrum and the quality of the fit is assessed by using a cost function comprised of two terms, the residual fit to the data and the deviation from the a priori atmospheric state (each weighted by their respective uncertainties). This cost function is minimised using a Levenburg-Marquardt iterative scheme~\citep{irwin08}.

\subsection{Reference atmosphere}
\label{sec:reference_atmosphere}

For this study, the Jovian atmosphere was divided into 39 levels between \SI{15}{\bar} and \SI{50}{\milli\bar}, equally spaced in log(p). The temperature and volume mixing ratio of each gas species is defined at each pressure level. 

The temperature profile is taken from Cassini CIRS observations~\citep{fletcher09}. The CIRS temperature profile is sensitive down to a pressure of \SI{800}{\milli\bar}, and has been extrapolated below this using a dry adiabat, consistent with the temperature profile found by the Galileo probe~\citep{seiff98} (see Figure~\ref{fig:weighting}b). The NH\textsubscript{3} and PH\textsubscript{3} profiles are also taken from~\citet{fletcher09} and have deep volume mixing ratios of $1.862 \times 10^{-4}$ and $1.86 \times 10^{-6}$ respectively.

The CH\textsubscript{4} profile was obtained from~\citet{nixon07} and has a deep volume mixing ratio of $1.81 \times 10^{-3}$. The CH\textsubscript{3}D profile assumes a constant ratio of CH\textsubscript{3}D/CH\textsubscript{4} of $8 \times 10^{-5}$~\citep{lellouch01}. The profile of H\textsubscript{2}O is poorly constrained. For this study, we assume a deep volume mixing ratio, fixed at $1 \times 10^{-3}$ ($\sim$ the solar abundance) with a constant relative humidity at higher altitudes. For the reference profile, this relative humidity is set to 10\%. The remaining minor atmospheric species, CO, GeH\textsubscript{4} and AsH\textsubscript{3} were assumed to be well mixed with volume mixing ratios of 1.0 ppb, 0.45 ppb and 0.24 ppb respectively~\citep{bezard02}.

\subsection{Line data}
\label{sec:line_data}

The line data sources that are used to produce the k-tables used in this study are described in~\citet{fletcher11}, along with the assumed broadening parameters and temperature dependences. This was updated to include information about additional GeH\textsubscript{4} isotopes obtained using STDS~\citep{wenger98}. As with the original GeH\textsubscript{4} isotope included in~\citet{fletcher11}, these isotopes were assumed to have a half-width of \SI{0.1}{\per\centi\meter}atm\textsuperscript{-1} and a temperature dependence of T\textsuperscript{0.75}.

As with~\citet{fletcher11}, lines were assumed to have a Voigt profile. The exception to this was NH\textsubscript{3}, which was updated to have a significantly sub-Lorentzian profile in the 5-\textmu m window, as found by~\citet{bailly04}.


\section{Sensitivity analysis}

\begin{figure}
\centering
\includegraphics[width=0.5\textwidth]{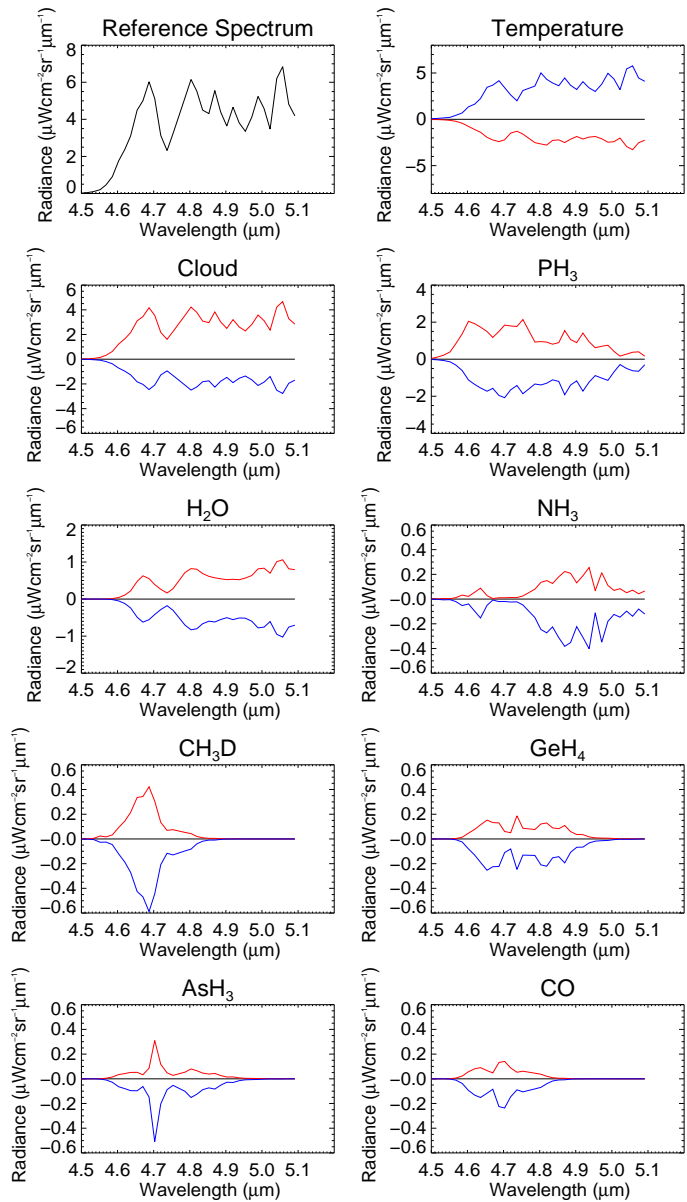}
\caption{The sensitivity of VIMS-resolution spectra to changes in different atmospheric parameters. The reference spectrum shows a synthetic spectrum computed for the atmospheric profile described in Section~\ref{sec:reference_atmosphere}, and the remaining panels show the (relative) changes to this spectrum that are caused by varying each parameter. The red (blue) lines correspond to a decrease (increase) of \SI{20}{\kelvin} for the temperature, a factor of 20\% decrease (increase) for the cloud opacity and a factor of 2 decrease (increase) for the gaseous species.}
\label{fig:sensitivity}
\end{figure}

The radiative transfer model described in Section~\ref{sec:radiativetransfer} was used to perform a sensitivity analysis, where each parameter was individually altered in order to observe its effect on the spectrum. Initially, a reference spectrum was produced using the atmospheric profile described in Section~\ref{sec:reference_atmosphere}. For this sensitivity analysis, a single, spectrally-flat cloud layer was inserted at \SI{0.8}{\bar} with an optical thickness of 10. The atmospheric paramaters were then individually altered, and the new spectra were compared to the reference spectrum. For the molecular species, the entire a priori profile was scaled by factors of 0.5 and 2.0. For water, the deep volume mixing ratio was held constant at the solar abundance and the relative humidity was varied by factors of 0.5 and 2.0. The cloud opacity was varied by factors of 0.8 and 1.2, and the temperature profile was shifted by $\pm$\SI{20}{\kelvin}. The results of this process can be seen in Figure~\ref{fig:sensitivity}, where the initial reference spetrum is shown, alongside nine frames showing the difference between the altered profiles and the reference. 

The gaseous species with the most significant impact on the shape of the VIMS-resolution 5-\textmu m spectrum are PH\textsubscript{3} and H\textsubscript{2}O. Although NH\textsubscript{3}, CH\textsubscript{3}D, GeH\textsubscript{4}, AsH\textsubscript{3} and CO all have spectral lines in this region, the relatively low spectral resolution means that they have a minor impact on the VIMS spectra. 

The single parameter that has the largest impact on the spectrum is the opacity of the cloud. Because the model cloud is spectrally flat and located well above the weighting function peak, the entire wavelength range is affected equally; changing the optical thickness scales the entire spectrum up or down, and a relatively minor change in the opacity has a significant impact on the average radiance. Shifting the temperature profile similarly scales the entire spectrum up or down, but to a lesser extent; to reproduce an effect which is comparable to a 20\% change of in cloud opacity, the temperature shift would have to be as large as \SI{20}{\kelvin}. A more realistic tropospheric temperature difference \SI{2}{\kelvin} between a moist and a dry adiabat~\citep{lewis95} produces a negligible effect on the spectrum when compared to the variability in cloud opacity. We therefore fix the temperature profile to the dry adiabat used in the reference profile, and note that this may produce a small error in the retrieved cloud opacities.

\begin{figure*}
\centering
\includegraphics[width=0.85\textwidth]{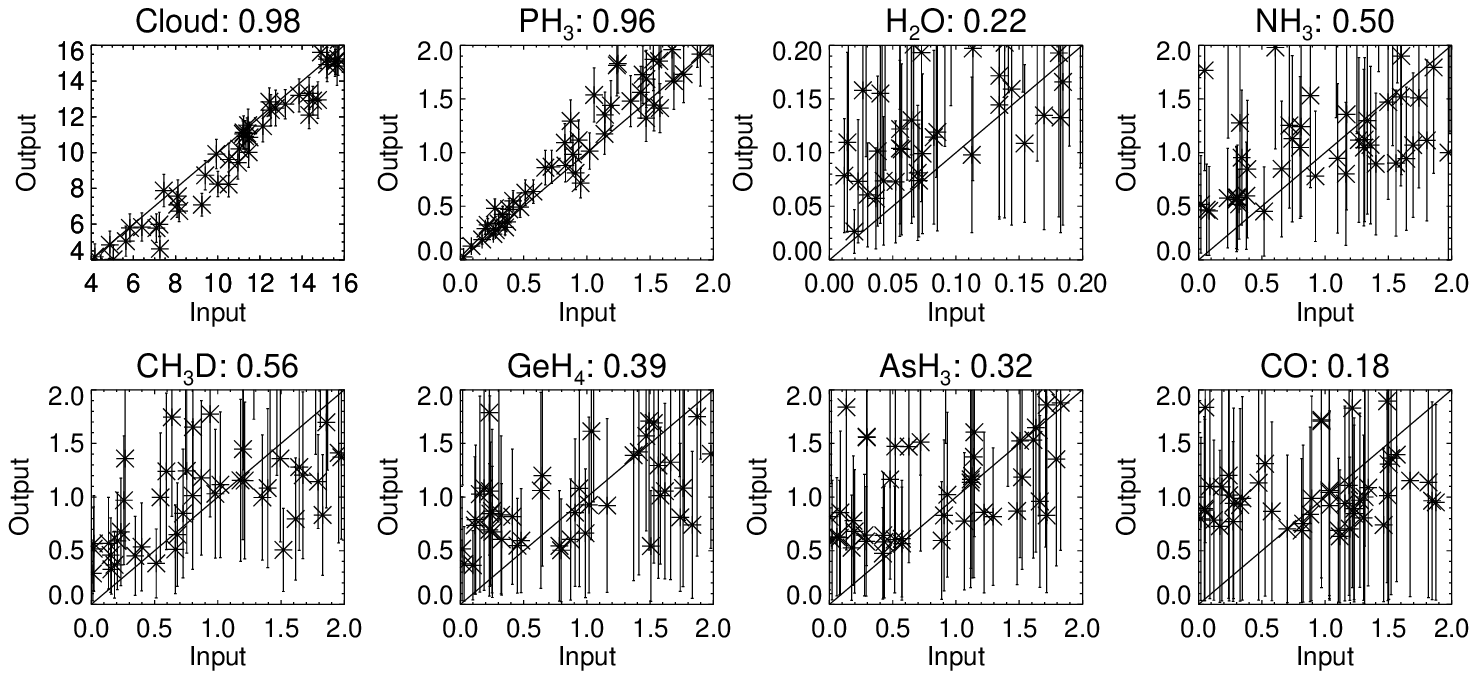}
\caption{The retrievability of different atmospheric paramaters for VIMS-resolution spectra. In each case, the true `input' value is plotted against the retrieved `output' value. For H\textsubscript{2}O, the input and output values correspond to the relative humidity, for the cloud, they correspond to cloud opacity and for the remaining gaseous species, they correspond to the scaling factors of the a priori profile. A perfect retrieval would result in the plotted points lying along the diagonal line. The Pearson product-moment correlation coefficients are given in the titles.}
\label{fig:real_vs_ret}
\end{figure*}

In addition to the degeneracy between cloud and temperature, there are further degeneracies  between gaseous species which complicate 5-\textmu m retrievals at this spectral resolution. The regions of sensitivity for each parameter are not unique, but instead overlap, leading to difficulties in disentangling the effects of each individual parameter to produce a reliable retrieval. This is further illustrated by Figure~\ref{fig:real_vs_ret}, which shows the results of a retrievability test. 75 radiative transfer models were run with random perturbations applied to the reference profile. Random noise was then added to these synthetic spectra, in accordance with the 12\% noise estimate for VIMS. The noisy spectra were then used as the input for NEMESIS retrievals. Figure~\ref{fig:real_vs_ret} compares output was compared with the `true' input values, and provides the Pearson product-moment correlation coefficient for each parameter. A correlation coeffiect of 1 indicates a perfect positive correlation, while a coefficient of 0 indicates that there is no linear relationship between the two variables.

The results tally with Figure~\ref{fig:sensitivity}; changes in cloud cover and the PH\textsubscript{3} volume mixing ratio have the most significant impact on the spectrum and can therefore be accurately retrieved, with Pearson product-moment correlation coefficients that are close to 1. In contrast, the retrievals for the other gaseous species are poor, and the retrieved values are only weakly correlated with the `true' input values. Based on these results, the abundances of NH\textsubscript{3}, CH\textsubscript{3}D, GeH\textsubscript{4}, AsH\textsubscript{3} and CO will be held fixed at their a priori values during the remainder of this study. Although the retrievals of H\textsubscript{2}O are fairly poor, we have no reliable a priori profile for water vapour, so the relative humidity of water will be allowed to vary, along with the volume mixing ratio of PH\textsubscript{3} and the cloud optical thickness.

%

\section{Basic retrievals}
\label{sec:basic}

Having explored the sensitivity of the 5-\textmu m region to different parameters and the potential limitations of any retrievals, we now turn to modelling the VIMS data itself using the NEMESIS retrieval algorithm. As with the sensitivity analysis, PH\textsubscript{3} and H\textsubscript{2}O were each allowed to vary via a single parameter. For PH\textsubscript{3} this was a scaling parameter that effectively altered the deep volume mixing ratio, and for H\textsubscript{2}O this was the relative humidity above a fixed deep volume mixing ratio. For each variable, large errors were assigned to a priori values. Testing was carried out to determine whether allowing additional degrees of freedom in these gases would improve the fit, but this did not make a difference in either case. 

As an initial model, we use a single, compact, 5-km thick cloud, located at \SI{0.8}{\bar}. The cloud base pressure is well above the weighting function peak (4-\SI{8}{\bar}) and is the approximate location of the predicted NH\textsubscript{3}-ice cloud. Figure~\ref{fig:spectra} shows that the shape of the spectra are almost identical at all latitudes, despite large variations in the absolute radiance. The cloud parameters therefore cannot vary strongly with wavelength, as otherwise any spectral features would be more evident in the cool, optically thick spectra than in the warm cloud-free spectra. In this initial model, we therefore use a spectrally flat cloud, as has been previously suggested by both ground-based and space-based studies, including~\citet{drossart82a},~\citet{bezard83},~\citet{bjoraker86a} and~\citet{roos-serote06}. The scattering parameters for this cloud were fixed at values that are broadly representative of moderately-sized NH\textsubscript{3} particles: single-scattering albedo, $\omega=0.9$ and asymmetry parameter, $g=0.8$. 

\begin{figure}
\centering
\includegraphics[width=0.45\textwidth]{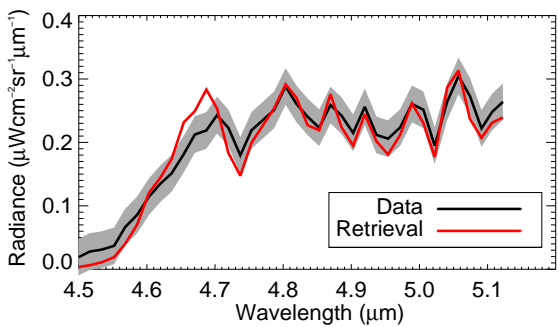}
\caption{Fit obtained using nightside data from the equatorial zone. The VIMS data is shown in black (with the error shown in grey) alongside the best-fit retrieved spectrum in red.}
\label{fig:retrieval_fit}
\end{figure}

An example of the results of these retrievals can be seen in Figure~\ref{fig:retrieval_fit}; we are able to produce a good fit to the VIMS 5-\textmu m data despite using a very simple cloud model and relatively few free parameters. The one exception to the otherwise good fit is the 4.65-\SI{4.75}{\micro\meter} region of the spectrum, where there is an apparent offset between the VIMS data and the retrieved spectrum. Many attempts were made to solve this issue, including allowing additional gaseous species to be retrieved, altering the parametrisations for the gases (including PH\textsubscript{3}), inserting multiple cloud decks and varying the vertical cloud profile. Similar issues can be seen in previous 5-\textmu m studies of Jupiter using Galileo/NIMS, including~\citet{roos-serote98} and~\citet{nixon01}.~\citet{roos-serote98} also had some difficulties in fitting a similar region of the high-resolution ISO/SWS spectra. The consistency of the problem suggests that the problem is with the models rather than the data. The mismatch is similar for both the cool and the warm spectra, suggesting that it is not due to a cloud spectral feature, and is more likely to be due to missing spectral lines from a gaseous species. As with~\citet{roos-serote98}, we exclude this part of the spectrum to ensure that the retrievals are not driven by an attempt to fit this one region.

%

\section{Cloud constraints}
\label{sec:cloud_constraints}

Instead of making assumptions about the composition of Jupiter's clouds, we sought to determine the range of cloud parameters that were consistent with the VIMS data. In Section~\ref{sec:basic}, we showed that a simple cloud model, consisting of a single, compact, spectrally flat cloud deck, is able to achieve a good fit to data. In this section, we further explore the cloud parameter space in order to constrain the cloud properties.

\subsection{Vertical structure}
\label{sec:vertical_structure}

\begin{figure*}
\centering
\includegraphics[width=0.65\textwidth]{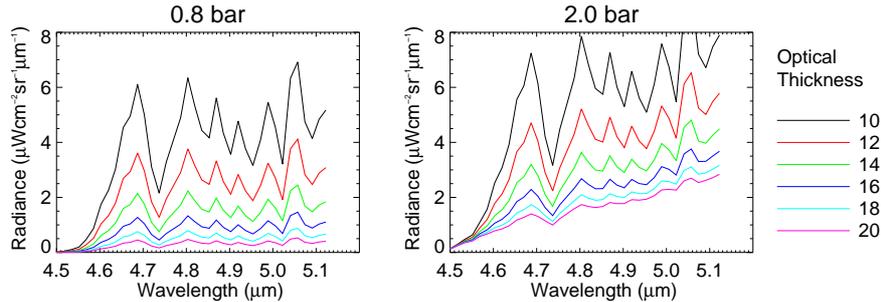}
\caption{Sensitivity of spectra to cloud locations. Each panel shows the spectrum for a variety of different cloud optical thicknesses. The upper panel corresponds to a compact cloud located at \SI{0.8}{\bar} and the lower panel corresponds to a compact cloud located at \SI{2.0}{\bar}.}
\label{fig:forward_cloud_locations}
\end{figure*}

Preliminary conclusions about the vertical location of the tropospheric cloud decks can be drawn by considering the VIMS spectra alone. As described in Section~\ref{sec:observations}, Figure~\ref{fig:spectra} shows that the shape of the spectra are almost identical at all latitudes, despite large variations in the absolute radiance. This immediately tells us that the main cloud decks must be located well above the 4-\SI{8}{\bar} region where the weighting functions peak (see Figure~\ref{fig:weighting}); otherwise, thick clouds would block out the radiation from some parts of spectrum more than others, giving a different shape to to the 5-\textmu m spectrum depending on the brightness temperature. Additionally, having an optically thick cloud at a deeper, hotter part of the atmosphere will also start to introduce a blackbody slope into the spectrum, which is not observed in any of the cool spectra. These effects can be seen in Figure~\ref{fig:forward_cloud_locations}, which shows the results of inserting a cloud at two different locations in Jupiter's troposphere and gradually increasing the optical thickness. In the upper panel, the cloud is located relatively high up in Jupiter's atmosphere, and increasing the optical thickness decreases the observed radiance, but maintains the same spectral shape, just as is observed in the VIMS spectra. In comparison, the lower panel shows the results for a cloud that has been placed deeper in the atmosphere, at \SI{2.0}{\bar}. Although the spectral shape remains similar for the lower opacities, the spectral shape starts to change for high optical thicknesses. In addition to changing the shape of the spectrum, this second case simply cannot reproduce the low average radiances observed in the cool zones in Figure~\ref{fig:spectra}; at these deeper pressures, the thermal emission from the cloud itself is higher and even an optically thick cloud therefore cannot reproduce the observed low radiances.

\begin{figure*}
\centering
\includegraphics[width=0.9\textwidth]{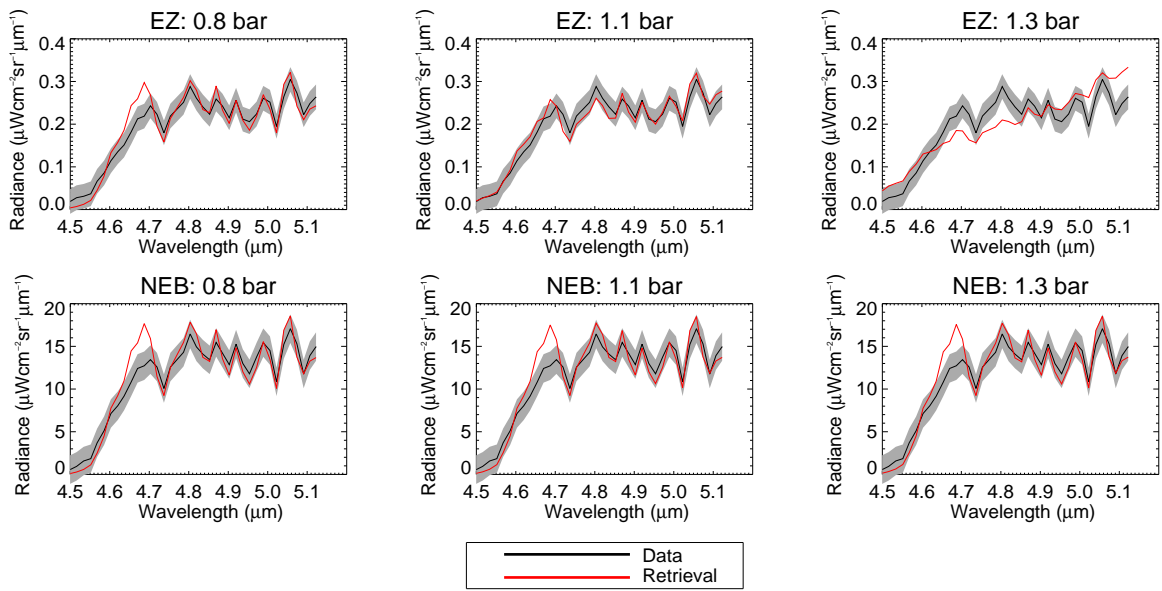}
\caption{Retrievals of VIMS spectra, with different assumptions about cloud location. Retrievals from the cool Equatorial Zone are shown at the top and retrievals from the warm North Equatorial Belt are on the bottom. Each column corresponds to a different cloud base location.}
\label{fig:cloud_retrieval}
\end{figure*}

The constraints on the vertical location of the clouds were further explored by running retrievals with a wide range of base pressures for the single, compact cloud. This was done for both a cool spectrum from the EZ and a warm spectrum from the NEB, and the results for two of the altitudes tested are shown in Figure~\ref{fig:cloud_retrieval}. When the cloud is located at \SI{0.8}{\bar}, good fits are obtained for both the EZ and the NEB. As the cloud is moved to progressively deeper pressures, the NEB fits initially remains good, but the EZ fits worsens significantly. The cut-off for a reasonable EZ fit is approximately \SI{1.2}{\bar}, while the NEB fits start to deteriorate for cloud base pressures of \SI{3.0}{\bar} or higher. The poor fit at deeper pressure levels is a result of the phenomena shown in Figure~\ref{fig:forward_cloud_locations}; as the cloud is moved deeper into the atmosphere, it starts to introduce a slope into the spectrum that is not seen in the data. In order to achieve a good fit for both warm and cool spectra using a single cloud deck, the clouds must be located at an altitude of \SI{1.2}{\bar} or higher.

In addition to exploring the location of the primary cloud opacity, testing was also carried out to investigate the vertical profile of the cloud. Additional cloud decks, with independently variable opacities, were included in the retrievals, but this made very little difference to the quality of the fits obtained. Similarly, extended cloud decks were used instead of vertically compact clouds, but this did not improve the fit either.

In summary, the VIMS 5-\textmu m nightside data can be fit using a simple cloud model, consisting of a single, compact, spectrally flat cloud deck whose opacity varies as a function of latitude, provided that this cloud deck is located at pressures less than \SI{1.2}{\bar}. It is important to note that this does not rule out more complicated cloud structures, including multiple cloud layers above \SI{1.2}{\bar}. The existence of deeper cloud decks is also possible, provided that the bulk of the 5-\textmu m opacity still originates from the upper clouds.

Similar analyses have previously been performed using observations from Galileo NIMS which cover the same wavelength range. ~\citet{roos-serote06} used a grey cloud model to fit the NIMS data, and found it had to be located above \SI{2.0}{\bar}. Our study provides a tighter constraint (p\textless \SI{2.0}{\bar}) because the global VIMS data includes spectra from very cold regions, while the NIMS study was restricted to warmer regions. It is the cooler spectra that provide the stronger constraint on the cloud location. Other NIMS studies have this same discrepency; ~\citet{irwin01}, ~\citet{nixon01} and ~\citet{irwin02} all place their principal cloud decks at pressures of 1-\SI{2}{\bar}, but since they only use warmer spectra, they are able to place the the clouds deeper in the atmosphere and still achieve good fits.

The 5-\textmu m VIMS results are consistent with Cassini observations made at other wavelengths. \citet{sromovsky10b} analysed the 3-\textmu m segment of the VIMS data. Using their four-layer multiple-scattering model, they found that the deepest, highest opacity cloud had a spatially variable cloud base that was located between 0.79 and \SI{1.27}{\bar}. Using a narrow spectral window from the CIRS instrument centered on \SI{7.18}{\micro\meter},~\citet{matcheva05} found that the cloud absorption coefficient peaks at 0.9-\SI{1.1}{\bar}. These locations are broadly consistent with the VIMS data, but we can neither confirm nor rule out the latitudinal variability in the cloud base pressures; we can achieve a good fit with a single cloud base pressure of \SI{1.2}{\bar} or lower, but a range of pressures is also possible, provided the cloud is located above \SI{3.0}{\bar} for the warm spectra from the belts and above \SI{1.2}{\bar} for the cool spectra from the zones.


\subsection{Spectral features}
\label{sec:spectral_features}

As described in Section~\ref{sec:basic}, the similar spectral shapes in Figure~\ref{fig:spectra} suggest that the clouds providing the majority of the 5-\textmu m opacity must be relatively spectrally flat, at least at the VIMS spectral resolution. Having established that a completely grey cloud is able to produce a good fit to the VIMS data, we now explore the extent to which spectral features are compatible which the observations.

\begin{figure}
\centering
\includegraphics[width=0.4\textwidth]{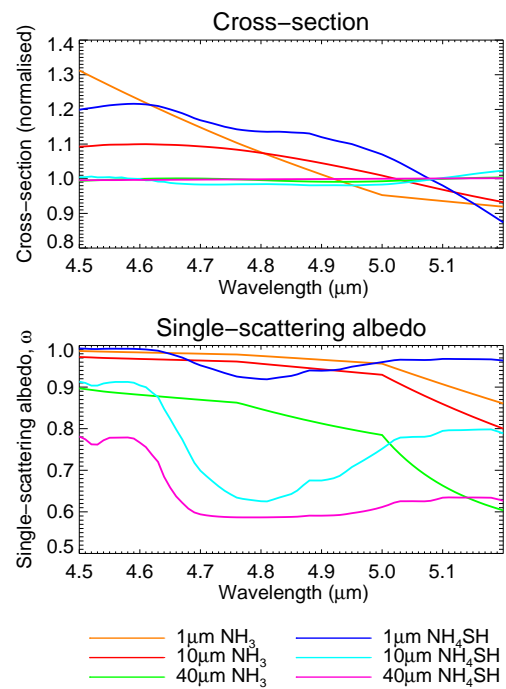}
\caption{Spectral parameters for NH\textsubscript{4}SH and NH\textsubscript{3} particles of different sizes, calculated via Mie theory.}
\label{fig:cloud_parameters}
\end{figure}

The two expected cloud materials in the middle troposphere are NH\textsubscript{3}-ice and NH\textsubscript{4}SH-ice. Figure~\ref{fig:cloud_parameters} shows the extinction cross-section and single-scattering albedo as a function of wavelength for a range of particle sizes. These functions have been calculated through Mie theory, assuming perfectly spherical particles. From this figure, it is clear that the scattering and absorption parameters of pure particles of NH\textsubscript{3} and NH\textsubscript{4}SH vary significantly with wavelength within the 5-\textmu m window. Larger particles have a relatively flat extinction cross-section, but a strongly wavelength-dependent single scattering albedo, while this is reversed for smaller particles. 

\begin{figure}
\centering
\includegraphics[width=0.4\textwidth]{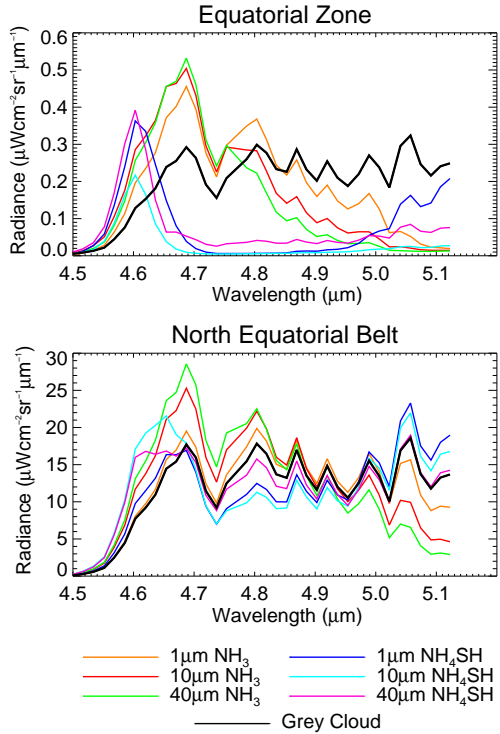}
\caption{The effect of different cloud particles on the shape of the spectra. The black line in each case gives the fit obtained using a spectrally flat cloud. The spectral features are more pronounced in the Equatorial Zone, where the clouds are optically thick.}
\label{fig:real_clouds}
\end{figure}

The effect of the wavelength-dependent cloud parameters on the spectral shape can be seen in Figure~\ref{fig:real_clouds}. In each panel, the black line shows the model fit to the VIMS data that is obtained using a spectrally flat cloud. The remaining lines show the effect of using a ``real'' cloud material, rather than a grey cloud. To make the impact of the clouds clear, the atmospheric composition (H\textsubscript{2}O and PH\textsubscript{3}) has been held constant for each cloud material, and the cloud opacity alone has been varied in order to produce an average radiance as close to the true value as possible. Performing a full retrieval does not however improve the fit sufficiently, and in some cases the algorithm attempts to retrieve an unphysically high abundance of the gaseous species in order to compensate for the poor cloud fit.

Figure~\ref{fig:real_clouds} clearly shows that these ``real'' clouds are unable to reproduce the shape of the VIMS spectra. This is particularly the case for Equatorial Zone, where the clouds are optically thick and any spectral features therefore become more prominent. The varying single-scattering albedo is the primary cause of these poor fits; the sharp gradient at $\sim$\SI{4.65}{\micro\meter} seen in Figure~\ref{fig:cloud_parameters} for the NH\textsubscript{4}SH particles translates into the sharp gradient seen in Figure~\ref{fig:real_clouds}, while the gradual slope for the NH\textsubscript{3} produces a slope in the spectra.

Despite being the most plausible materials for the thick tropospheric clouds, pure NH\textsubscript{3}-ice and pure NH\textsubscript{4}SH-ice are not consistent with the VIMS 5-\textmu m observations. However, there are several possible explanations for this. The presence of several cloud decks made up of particles of different materials and/or sizes can act together to ``blur out'' the individual spectral features, giving a net effect of a roughly grey cloud. This effect, along with the fact that the study focussed on warmer regions where the clouds are thinner, allowed~\citet{irwin98} to fit the Galileo NIMS 5-\textmu m spectra using a 4-level cloud structure, made up of \SI{0.5}{\micro\meter} tholins, \SI{0.75}{\micro\meter} NH\textsubscript{3} particles, and \SI{0.45}{\micro\meter} and \SI{50}{\micro\meter} NH\textsubscript{4}SH particles.

Alternatively, the clouds could be made of a different material altogether, or be masked by deposits from another material which causes the cloud to be spectrally flat. Coating of ammonia particles by other substances has been suggested by~\citet{baines02} as an explanation for the absence of spectroscopically identifiable ammonia clouds in across the majority of the planet and ~\citet{kalogerakis08} found that thin layers of hydrocarbons are able to alter the spectral features at 3 and \SI{9}{\micro\meter}.

The 5-\textmu m data alone cannot distinguish between these various possible scenarios, but the net effect in each case is a roughly grey cloud. We choose the simplest option for reproducing this, and continue the analysis using a single spectrally-flat cloud.


\subsection{Scattering properties}

The choice of scattering parameters affects the thermal scattering on the planet's nightside. We now explore the range of scattering parameters that are consistent with the VIMS nightside observations. Using the nightside nadir data described in Section~\ref{sec:data_selection} and used in Sections~\ref{sec:vertical_structure} and~\ref{sec:spectral_features}, retrievals were performed using many values of $\omega$ and $g$, covering the full range of physically realistic particles. Although the numerical results of the retrievals varied with these different values (a higher single-scattering albedo requires a higher cloud optical thickness for the same spectrum), the fits produced were very uniform. Many different cloud parameters were capable of reproducing the nadir VIMS spectra, so the values of $\omega$ and $g$ cannot be constrained from the nadir data alone.

Additional insights can be gained by considering observations made at a range of emission angles. In the Jovian troposphere, temperature decreases with altitude, leading to limb darkening: the radiance observed decreases towards the edge of the disc. The extent of this limb darkening is heavily dependent on the cloud decks that are present in the troposphere; by comparing the spectra at different emission angles, we can constrain the scattering parameters of these clouds.

\begin{figure}
\centering
\includegraphics[width=0.4\textwidth]{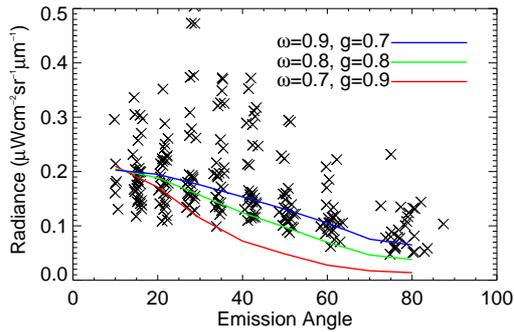}
\caption{Limb darkening for different cloud parameters. The scatter plots show the VIMS data as a function of emission angle, averaged over the 5-\textmu m spectral window. The large spread is due to the spatial inhomogeneity of the clouds, even within a single latitude band. Overplotted are the synthetic limb darkening curves for three example sets of cloud parameters (see Figure~\ref{fig:limb_contour} for the full range of parameters tested).}
\label{fig:limb}
\end{figure}

The relatively low spatial resolution of the VIMS data means that it is not possible to isolate a single atmospheric feature and compare its appearance at different emission angles. Instead, the scatter plots in Figure~\ref{fig:limb} show the limb darkening observed in the equatorial region of the planet ($-2.5^{\circ}$ to $2.5^{\circ}$). The equatorial region was chosen as it provides the largest range of emission angles, and exhibits less spatial inhomogeneity than other parts of the planet. Each point corresponds to the mean 4.5-\SI{5.2}{\micro\meter} radiance from a single pixel. 

The large spread of points is due to the inhomogeneity of the cloud thickness, even within a single latitudinal band (as seen in Figure~\ref{fig:vimsmap}). Nevertheless, the general limb darkening trend can be seen - as the emission angles increases towards the limb, the radiance decreases.

The broad scatter of points in Figure~\ref{fig:limb} means that a simultaneous retrieval of data at different emission angles is unlikely to be productive. Two points at different positions along the disc are likely to have very different cloud opacities, in addition to having different emission angles. However forward modelling does allow us to place some constraints on the cloud parameters that are consistent with the observed data.

For a range of cloud parameter combinations (single-scattering albedo, $\omega$ and asymmetry parameter, $g$), retrievals were performed using nadir (low-emission angle) data. The retrieved parameters from the nadir data were then used to forward-model spectra corresponding to higher emission angles. The results of this process for three example combinations are shown by the solid lines in Figure~\ref{fig:limb}.

Each limb darkening curve is anchored to the same point because of the initial retrieval, but they all vary differently with emission angle. Despite the broad scatter of points, it is immediately apparent from Figure~\ref{fig:limb}a that the case $\omega=0.9, g=0.7$ is consistent with the data, but $\omega=0.7, g=0.9$ is not.

\begin{figure}
\centering
\includegraphics[width=0.42\textwidth]{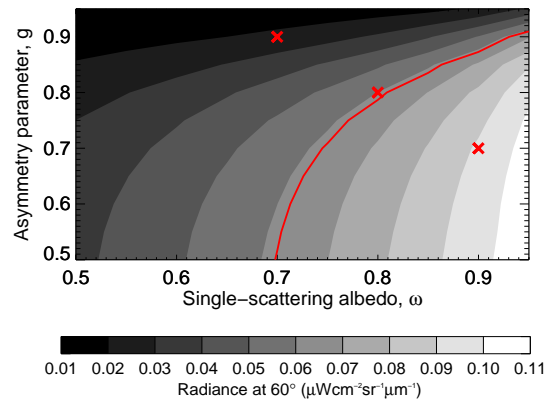}
\caption{The synthetic curve radiances at an angle of $60^{\circ}$, for a range of scattering parameters. The red crosses mark the three example parameters used in Figure~\ref{fig:limb}. The red line marks the minimum observed radiance from the VIMS data; scattering parameters above this line are not consistent with the data. Cases where $\omega$ is low and $g$ is high are ruled out.}
\label{fig:limb_contour}
\end{figure}

This retrieval and forward-modelling process was repeated for more parameter combinations. For each case, the radiance at an emission angle of $60^{\circ}$ was recorded and is plotted in Figure~\ref{fig:limb_contour}. The red line corresponds to a radiance of \SI{0.062}{\micro\watt\per\centi\meter\per\steradian\per\micro\meter}, the lowest observed radiance at $60^{\circ}$ in the VIMS equatorial data. Scattering parameters which produce radiances larger than this value are consistent with the data; scattering parameters which produce smaller radiances are not.  Looking at Figure~\ref{fig:limb}, the blue curve falls within the acceptable fit region, the green curve is on the borderline, and the red curve does not give an acceptable fit. While a fairly large range of $\omega$ and $g$ values give an acceptable fit to the VIMS limb darkening data (the bottom-right segment of the plot), this figure rules out the case where $\omega$ is low and $g$ is high.

For this analysis, the location of the cloud deck was held constant at~\SI{0.8}{\bar}. The effect of varying the altitude of the cloud deck was investigated, and it was found that no additional constraint was provided on top of the results from Section~\ref{sec:vertical_structure}, i.e. the clouds can be located anywhere above \SI{1.2}{\bar}. The conclusions drawn from Figure~\ref{fig:limb_contour} are independent of the cloud location, provided that it is above the \SI{1.2}{\bar} level.

This is in good agreement with the results from~\citet{roos-serote06}, who performed a limb darkening analysis using pairs of \SI{5}{\micro\meter} data cubes from Galileo NIMS. With $g$ fixed at 0.8, they found that an optimum fit was obtained with $\omega=0.9\pm0.5$.

%
\begin{figure*}
\centering
\includegraphics[width=1.0\textwidth]{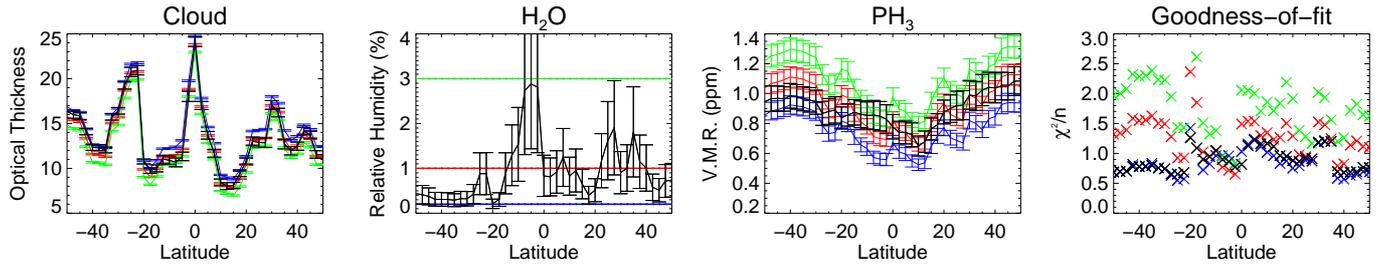}
\caption{North-south latitudinal retrievals of the cloud opacity, the H\textsubscript{2}O relative humidity and the PH\textsubscript{3} abundance from Jupiter's nightside. The different colours correspond to the four different conditions placed on the H\textsubscript{2}O relative humidity: allowed to vary (black), fixed at 0.2\% (blue), fixed at 1\% (red), fixed at 3\% (green). Along with the retrieved quantities is the goodness-of-fit as a function of latitude for each condition.}
\label{fig:latitudinal_retrieval}
\end{figure*}

\begin{figure}
\centering
\begin{subfigure}{0.5\textwidth}
\centering
\includegraphics[width=0.95\textwidth]{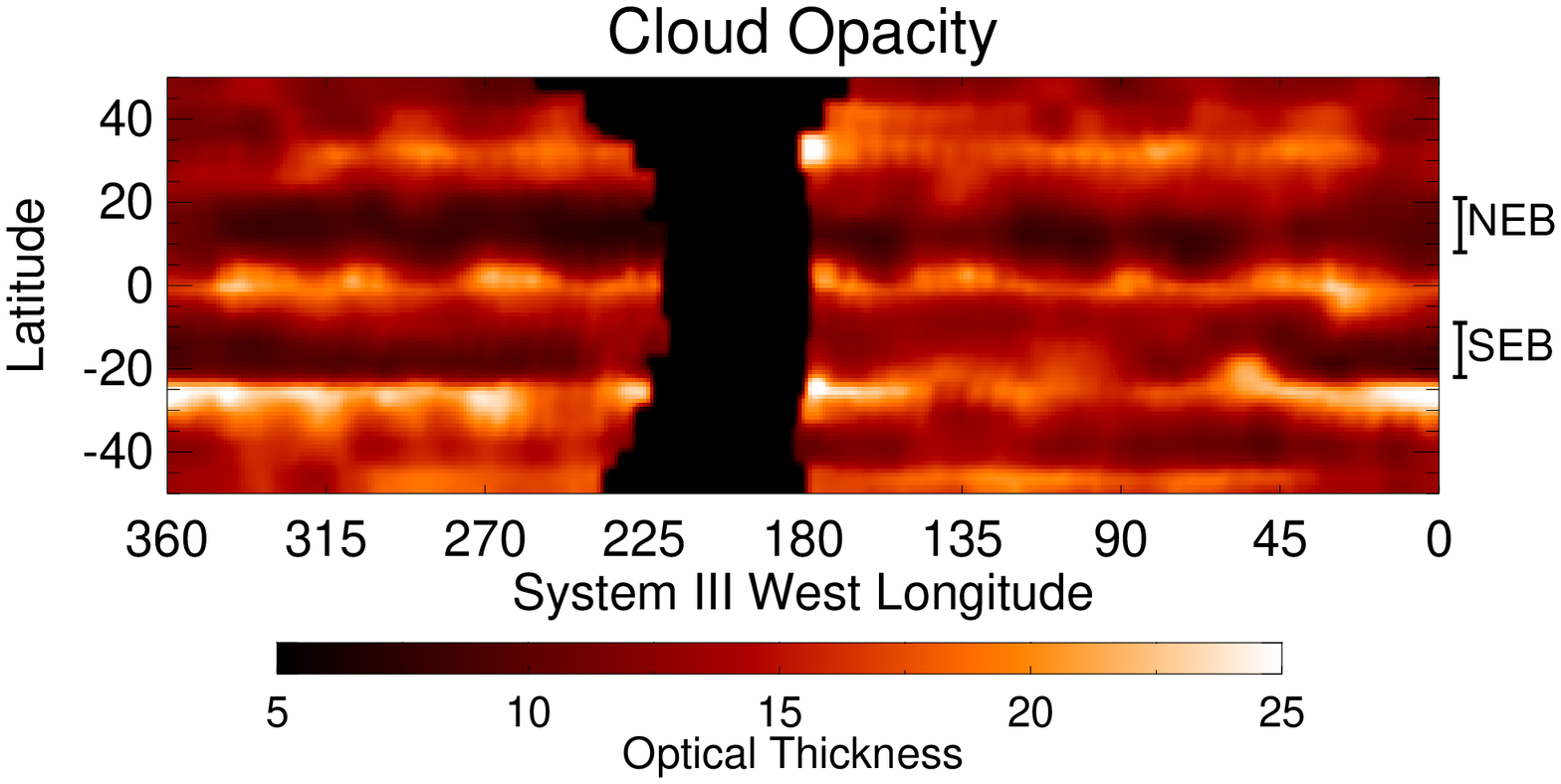}. 
\end{subfigure}
\begin{subfigure}{0.5\textwidth}
\centering
\includegraphics[width=0.95\textwidth]{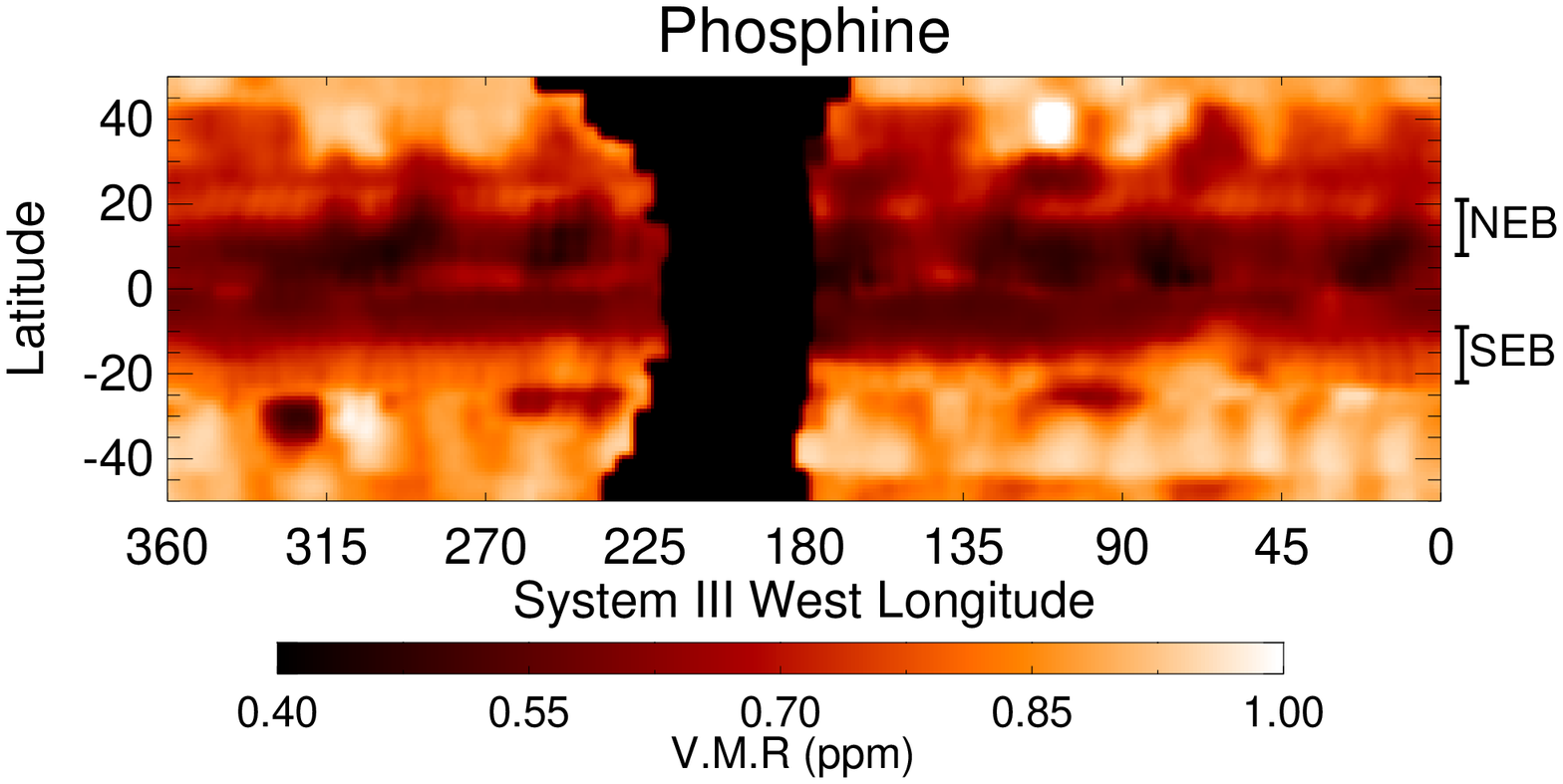}
\end{subfigure}
\caption{Global maps of the retrieved cloud opacity and PH\textsubscript{3} abundance from the VIMS nightside observations. The H\textsubscript{2}O relative humidity has been held fixed at 0.2\%. Smoothing has been applied to these maps.}
\label{fig:vimsmap_retrievals}
\end{figure}

\section{Global retrievals}
\label{sec:variations}

Using the best-fit cloud model from the previous section (a single, compact grey cloud located above \SI{1.2}{bar}), retrievals were run across the entire planet in order to investigate variability in the atmospheric parameters. As before, the free parameters in the retrievals were the cloud opacity, the water vapour relative humidity, and the scaling factor of phosphine (effectively the deep volume mixing ratio). Retrievals were run for each of the the forty-one $5^{\circ}$ latitudinal bins described in Section~\ref{sec:data_selection} and the results are shown in Figure~\ref{fig:latitudinal_retrieval}. The initial error estimate of 12\% lead to underfitting the data, so the errors were reduced to 6\% in order to give a reduced chi-squared value of $\sim$1 - the errors shown in grey are the formal errors on the retrieval, determined using the 6\% error value. 

\subsection{Water vapour}

For the initial retrieval, the H\textsubscript{2}O abundance was included as a free parameter. The results for this process are shown in black in Figure~\ref{fig:latitudinal_retrieval}. If allowed to vary, the zonally-averaged relative humidity ranges from $\sim$0.2\% to $\sim$3\%, with a maximum around the equator and a somewhat asymmetric appearance. However, Figure~\ref{fig:real_vs_ret} previously showed that H\textsubscript{2}O cannot be accurately retrieved due to extreme degeneracy and correlation with the other parameters, so these results should be taken with caution. 

To further test these results, we then fixed the H\textsubscript{2}O relative humidity at various different levels and re-ran the latitudinal retrievals. Figure~\ref{fig:latitudinal_retrieval} shows the results for three of these relative humidities: 0.2\% (blue), 1\% (red) and 3\% (green). The fourth panel of this figure shows the impact that each fixed value has on the goodness-of-fit. Although the higher fixed relative humidities (1\% and 3\%) lead to significantly worse fits at certain latitudes, the lowest value (0.2\%) produces fits that are comparable to the free retrieval. We note that removing water vapour entirely from the retrievals significantly worsened the fit.

For certain fixed H\textsubscript{2}O relative humidities, the cloud opacity and PH\textsubscript{3} abundance are able to adjust and provide an equally good fit to the data. By considering the changes in the chi-square parameter, we estimate that the maximum fixed relative humidity allowed is approximately 0.5\%. Since we are able to achieve a good fit using a single H\textsubscript{2}O profile at all latitudes (provided that the relative humidity is less than 0.5\%) we conclude that the VIMS 5-\textmu m spectra do not provide any evidence for latitudinal variability in water vapour. Previous studies using NIMS data~\citep{roos-serote00} have, however, suggested that there is considerable local variability in the H\textsubscript{2}O humidity; since these NIMS analyses made use of high spatial resolution observations of small regions of the planet, this discrepancy may be due to sub-pixel and zonal inhomogeneity in the relatively low resolution VIMS observations. Any small regions of elevated water may be rendered invisible by averaging over the large areas covered by the VIMS pixels, although a thorough exploration of gas and cloud degeneracies in the NIMS dataset is required to confirm this hypothesis.

Assessments of spatial variability of tropospheric H\textsubscript{2}O must await higher spectral resolution space-based measurement to better distinguish between the competing effects of water, cloud opacity and phosphine. 

\subsection{Phosphine}

The latitudinal distributions of PH\textsubscript{3} for the different H\textsubscript{2}O conditions are shown in Figure~\ref{fig:latitudinal_retrieval}, and Figure~\ref{fig:vimsmap_retrievals} shows the global distribution for the case where the H\textsubscript{2}O relative humidity is fixed at 0.2\%. These global maps were produced by binning the spectra into $10^{\circ} \times 10^{\circ}$ size bins and running a retrieval at each latitude and longitude point.

The different water profiles in Figure~\ref{fig:latitudinal_retrieval} lead to the following global averages for the PH\textsubscript{3} deep volume mixing ratio: 0.90$\pm$0.09 ppm (allowing H\textsubscript{2}O to vary), 0.76$\pm$0.05 ppm (0.2\% relative humidity), 0.92$\pm$0.6 ppm (1\% relative humidity), 1.09$\pm$0.06 ppm (3\% relative humidity). The best fits therefore suggest a deep volume mixing ratio of 0.76-0.90 ppm. These retrieved global averages are consistent with previous 5-\textmu m studies, including~\citet{bjoraker86a} who gives a value of 0.7 ppm from airborne spectroscopic observations, and~\citet{irwin98} who gives a value of 0.77 ppm from Galileo NIMS. However, analyses of Cassini CIRS 7.7-\SI{16.6}{\micro\meter} observations have reported higher values:~\citet{irwin04} fit the data using a deep volume mixing ratio that varies between 1.0 and 1.5 ppm, while~\citet{fletcher09} found that values of 1.8-1.9 ppm produced the best fit. Since CIRS is sensitive to higher altitudes than NIMS and VIMS, and we expect the phosphine abundance to decrease with altitude rather than increase, it is likely that this discrepancy is due to an inconsistency in the database line strengths between the two spectral regions.

Although the different water vapour profiles shown in Figure~\ref{fig:latitudinal_retrieval} lead to slightly different PH\textsubscript{3} retrievals, they each produce a similar latitudinal pattern: an enhanced abundance at high latitudes compared to low latitudes. This is a phenomenon previously noted in the northern hemisphere by ~\citet{drossart90}, who used high-resolution 5-\textmu m spectra to detect a 60\% enhancement of PH\textsubscript{3} at high northern latitudes compared to the NEB; our retrievals give an enhancement of $\sim$60-75\%. \citet{drossart90} suggest that this high-latitude enhancement could be either due to an increase in photochemical dissociation or a decrease in the eddy diffusion coefficient at high latitudes. 

\subsection{Clouds}

Figures~\ref{fig:vimsmap} and~\ref{fig:spectra} both showed that there was significant variation in 5-\textmu m radiance with latitude; as expected, Figures~\ref{fig:latitudinal_retrieval} and~\ref{fig:vimsmap_retrievals} shows that this is primarily due to variable cloud thickness. The cloud opacity retrieval shown in Figure~\ref{fig:vimsmap_retrievals} is the inverse of Figure~\ref{fig:vimsmap_night}; where the brightness temperature is high, the opacity is low, and vice versa. The absolute retrieved values of the cloud optical thickness are highly dependent on the chosen cloud parameters (and slightly dependent on the H\textsubscript{2}O relative humidity), but the relative changes are always reproduced, with high optical thicknesses in the zones ($0^{\circ}$,$\pm30^{\circ}$) and low optical thicknesses in the belts ($\pm15^{\circ}$).

Figure~\ref{fig:vimsmap} also showed that there is considerable variability within latitude bands. One region that has significant longitudinal variation is the SEB, where the brightness temperature varies with distance from the Great Red Spot. Figure~\ref{fig:vimsmap_retrievals} shows that this is primarily due to cloud opacity. The thickest clouds in the SEB occur to the west of the GRS ($50^{\circ}$W, $20^{\circ}$S), in its turbulent wake. It is not until the opposite side of the planet ($\sim225^{\circ}$W), that these clouds thin out and the brightness temperature increases. This phenomenon may be explained by turbulence dredging up material from deeper pressures, causing clouds to form. As the distance from the GRS increases, the atmosphere becomes more quiescent and we return to the ordinary, relatively cloud-free appearance of a belt.

%

\section{Reflected sunlight analysis}
\label{sec:dayside}

\begin{figure}
\centering
\includegraphics[width=0.45\textwidth]{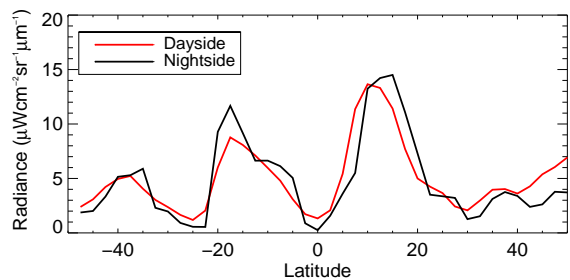}
\caption{Zonally averaged \SI{5.0}{\micro\meter} radiance as a function of latitude for both the nightside and the dayside.}
\label{fig:dayside_nightside}
\end{figure}

In addition to the nightside data, VIMS also made measurements on Jupiter's dayside. The majority of this study has focussed on the nightside data, as the absence of reflected sunlight considerably simplifies the analysis. In this final section, we seek to determine whether the results obtained from the nightside are consistent with the dayside observations.

Figure~\ref{fig:vimsmap} compared the global maps from the nightside and the dayside, and we noted that the contrast between the belts and the zones is smaller on the dayside than on the nightside. This can also be seen in Figure~\ref{fig:dayside_nightside}, which gives the zonal averages as a function of latitude. The additional component of reflected sunlight makes little difference in the warm belts and there are even points where the nightside is brighter than the dayside. This is because the belts are relatively cloud-free, so there is less reflection of sunlight. If we were looking at precisely the same cloud at different solar zenith angles, then we would expect a small increase due to reflected sunlight; the fact that we sometimes see a small decrease is due to the fact that the gap of several weeks between the measurements and the low spatial resolution means that we are not observing identical atmospheric conditions.

In the cooler, cloudier regions of the planet, such as the equator, the reflected sunlight component becomes more significant. The thicker clouds lead to more reflection, and ensure that the dayside is consistently brighter than the nightside. At the equator, the zonal average radiance at \SI{5.0}{\micro\meter} increases from \SI{0.25}{\micro\watt\per\centi\meter\per\steradian\per\micro\meter} to \SI{1.33}{\micro\watt\per\centi\meter\per\steradian\per\micro\meter}, an increase of more than 500\%. Again, part of this difference may be due to the fact that we are not necessarily observing identical regions of the planet, but reflected sunlight clearly accounts for a significant part of the dayside radiance from the zones. This flux difference between the nightside and dayside observations of the equatorial  zone is consistent with the analysis of~\citet{drossart98}, who studied the solar reflected component of Jupiter's 5-\textmu m spectra from Galileo NIMS and found that the minimum flux level was six times greater on the dayside than on the nightside. 

\begin{figure}
\centering
\includegraphics[width=0.45\textwidth]{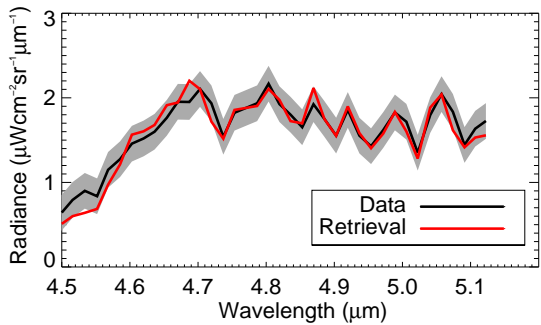}
\caption{Example of the fit obtained using dayside data from the equatorial zone.}
\label{fig:retrieval_day}
\end{figure}

We ran retrievals across the range of latitudes on the dayside, using the set of parameters from the previous sections: the volume mixing ratio of PH\textsubscript{3}, the relative humidity of water and the opacity of a single, compact, grey cloud located at~\SI{0.8}{\bar}. For these dayside retrievals, the reflected solar component is included in the radiative transfer model. We found that we were able to obtain a good fit to the data using this simple cloud model; an example of the fit obtained in the equatorial region is shown in Figure~\ref{fig:retrieval_day}.

\begin{figure*}
\centering
\includegraphics[width=1.0\textwidth]{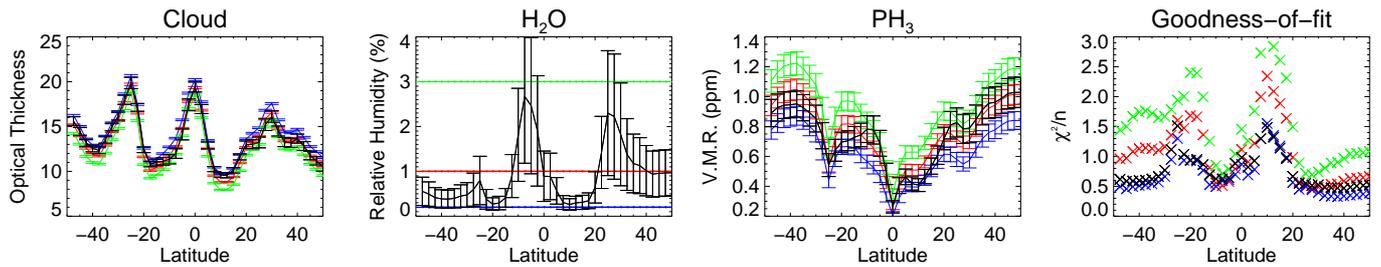}
\caption{North-south latitudinal retrievals of the cloud opacity, the H\textsubscript{2}O relative humidity and the PH\textsubscript{3} abundance from Jupiter's dayside. The different colours correspond to the four different conditions placed on the H\textsubscript{2}O relative humidity: allowed to vary (black), fixed at 0.2\% (blue), fixed at 1\% (red), fixed at 3\% (green). Along with the retrieved quantities is the goodness-of-fit as a function of latitude for each condition.}
\label{fig:latitudinal_retrieval_day}
\end{figure*}

The full results of the zonally-averaged retrievals are shown in Figure~\ref{fig:latitudinal_retrieval_day}. As with Figure~\ref{fig:latitudinal_retrieval}, the different colours refer to the different conditions placed on the H\textsubscript{2}O profile. The overall shapes of the retrievals are similar to the equivalent nightside plot (Figure~\ref{fig:latitudinal_retrieval}); the cloud opacities peak at the same latitudes, and the PH\textsubscript{3} abundance still exhibits an enhancement at high latitudes. There are, however, a few differences. Firstly, the retrieved values of PH\textsubscript{3} exhibit additional minima which correspond to the peaks in cloud opacity. We believe the addition of reflected sunlight accentuates the degeneracy between the cloud parameters and the PH\textsubscript{3} abundance, as it is in the PH\textsubscript{3} absorption wing at the short-wavelength edge of the window where the reflected sunlight component is most significant. A small change in the cloud scattering parameters can lead to dramatic changes in the retrieved PH\textsubscript{3} values (a phenomenon not seen on the nightside), so retrievals are unreliable. Higher resolution dayside spectroscopy would reduce this degeneracy, allowing more reliable PH\textsubscript{3} retrievals.

Secondly, although the cloud opacity has a very similar shape, the absolute values are slightly different. This is likely to be due to a combination of two factors: the opacities may be genuinely different, due to averaging over spatially inhomogeneous latitudinal band, and the retrievals may be mismatched due to slightly incorrect cloud scattering properties leading to an imbalance between thermal radiation and reflected sunlight. A more comprehensive study, taking into account shorter wavelengths, would be required to jointly constrain the cloud properties.

\section{Conclusions}

This paper uses the 2000/2001 observations of Jupiter made by the Cassini VIMS instrument in the 4.5-\SI{5.1}{\micro\meter} range to study the planet's tropospheric composition and cloud structure. This builds on previous work using the Galileo NIMS and Voyager IRIS datasets by making use of (i) the full global coverage afforded by VIMS, covering both warm and cool regions of the planet; (ii) the combination of nightside and dayside observations. We conclude that:

\begin{enumerate}

\item VIMS 5-\textmu m data (both nightside and dayside) can be modelled using a very simple cloud model, consisting of a single, compact, spectrally flat cloud.

\item The bulk of the 5-\textmu m opacity must be located sufficiently high in the troposphere that is does not impact the shape of the spectrum. For the coolest regions on the planet, with the thickest clouds, this requirement constrains the clouds to altitudes of \SI{1.2}{\bar} or higher. In warmer regions, the clouds can be placed deeper in the atmosphere and still achieve a good fit to the data.

\item The spectra from both cloudy and relatively cloud-free regions have very similar spectral shapes, so the clouds must be relatively spectrally-flat. Pure NH\textsubscript{3}-ice and pure NH\textsubscript{4}SH do not have sufficiently grey spectra, and are therefore inconsistent with VIMS data. It may be that spectral features in the clouds are masked by coating layers, or that multiple cloud decks act together to blur out any features.

\item The relative lack of limb darkening means that the cloud particles must be highly scattering. There is a degeneracy between the single-scattering albedo and the asymmetry parameter, but cases with a low single-scattering albedo and high asymmetry parameter are ruled out.

\item The majority of the 5-\textmu m global inhomogeneity can be accounted for by variations in the cloud opacity, with thick clouds in the zones and relatively cloud-free belts.

\item The retrieved globally-averaged deep volume mixing ratio for PH\textsubscript{3} was 0.76$\pm$0.05 ppm (with the H\textsubscript{2}O relative humidity fixed at 0.2\%), consistent with previous 5-\textmu m studies. The latitudinal retrieval of PH\textsubscript{3} shows an enhancement in the abundance at high latitudes.

\item The VIMS 5-\textmu m spectra do not provide any evidence for latitudinal variability in the H\textsubscript{2}O relative humidity; if the H\textsubscript{2}O abundance is held fixed, the cloud opacity and PH\textsubscript{3} abundance are able to adjust in order to produce an equally good fit to the data. The low spectral resolution and high degeneracy between gases mean that H\textsubscript{2}O cannot be accurately retrieved, but a fixed relative humidity of less than 0.5\% is able to provide a good fit to the data at all latitudes. 

\item The low spectral resolution of VIMS also prevents the accurate retrieval of other gaseous species with weaker signatures in the 5-\textmu m window: NH\textsubscript{3}, CH\textsubscript{3}D, GeH\textsubscript{4}, AsH\textsubscript{3} \& CO. High resolution spectroscopy is required to retrieve abundances of these species and to search for any spatial variability.

\end{enumerate}

\section*{Acknowledgements}

Giles was supported via a Royal Society studentship, and Fletcher was supported via a Royal Society Research Fellowship at the University of Oxford. Irwin acknowledges the support of the United Kingdom Science and Technology Facilities Council. We thank the Cassini/VIMS team for planning and implementing these observations.


\bibliography{master}


\end{document}